\documentclass[final,5p,times,twocolumn,authoryear]{elsarticle}
\usepackage[hidelinks]{hyperref}
\usepackage{lineno}
\usepackage{import}
\hypersetup{colorlinks=true, linkcolor=blue, filecolor=magenta, urlcolor=cyan}
\usepackage{amsmath}
\usepackage{tabularx}
\usepackage[T1]{fontenc}
\usepackage[utf8]{inputenc}
\usepackage{lmodern}
\usepackage{graphicx}
\usepackage{booktabs}
\usepackage{amsmath, amssymb}
\usepackage{siunitx}
\usepackage{caption}
\usepackage{enumitem}
\usepackage{hyperref}
\usepackage{titlesec}
\usepackage{endnotes} 
\let\footnote=\endnote
\usepackage{float}
\usepackage{textgreek}
\usepackage{amssymb}
\usepackage{chngcntr}

\journal{Elsevier}

\begin{document}
\begin{frontmatter}
\title {Association of Progressive PPFE and Mortality in Lung Cancer Screening Cohorts}

\address[a]{Satsuma Lab, Hawkes Institute, University College London, London, UK}
\address[b]{Department of Respiratory Medicine, University College London, London, UK}
\address[c]{Department of Computer Science, University College London, London, UK}
\address[d]{Guy’s Cancer Centre, Guy’s and St Thomas’ NHS Foundation Trust, London, UK}
\address[e]{Radiology Department, Hospital de la Santa Creu i Sant Pau, Sant Quintí, Barcelona, Spain}
\address[f]{Department of Medicine, University of Iceland, Iceland}
\address[g]{NIHR Manchester Biomedical Research Centre, Manchester Academic Health Science Centre, Central Manchester University Hospitals, NHS Foundation Trust, Manchester, UK}
\address[h]{Division of Informatics, Imaging and Data Sciences, School of Health Sciences, University of Manchester, Manchester, UK}
\address[i]{UCL Hawkes Institute, University College London, London, UK}
\address[j]{Lungs for Living, Department of Respiratory Medicine, University College London, London, UK}

\author[a,b,c]{Shahab Aslani\corref{corr1}}
\author[a,c]{Mehran Azimbagirad}
\author[a,b]{Daryl Cheng}
\author[a]{Daisuke Yamada}
\author[a]{Ryoko Egashira}
\author[a,c]{Adam Szmul}
\author[a]{Justine Chan-Fook}
\author[a,b]{Robert Chapman}
\author[a,d]{Alfred Chung Pui So}
\author[a]{Shanshan Wang}
\author[a]{John McCabe}
\author[a]{Tianqi Yang}
\author[e]{Jose M Brenes}
\author[f]{Eyjolfur Gudmundsson}
\author[]{The SUMMIT Consortium\corref{corr2}}
\author[g,h]{Susan M. Astley}
\author[c,i]{Daniel C. Alexander}
\author[j]{Sam M. Janes}
\author[a,b]{Joseph Jacob}
\cortext[corr1]{Corresponding author: a.shahab@ucl.ac.uk}
\cortext[corr2]{Summit Consortium authors and affiliations listed at end of file}

\begin{abstract}

\textbf{Background:} Pleuroparenchymal fibroelastosis (PPFE) is an upper lobe predominant fibrotic lung abnormality associated with increased mortality in established interstitial lung disease. However, the clinical significance of radiologic PPFE progression in lung cancer screening (LCS) populations remains unclear. We investigated whether progressive PPFE quantified on low-dose CT scans independently associates with mortality and respiratory morbidity in two lung cancer screening cohorts.\\

\textbf{Methods:} We analysed longitudinal low-dose CT scans and clinical data from two LCS studies: National Lung Screening Trial (NLST; n=7,980); SUMMIT study (n=8,561). An automated algorithm quantified PPFE volume on baseline and follow-up scans. Annualised change in PPFE was derived and dichotomised using a distribution-based threshold to define progressive PPFE. Associations between progressive PPFE and mortality were evaluated using Cox proportional hazards models adjusted for demographic and clinical variables. In SUMMIT cohort, associations between progressive PPFE and clinical outcomes were assessed using incidence rate ratios (IRR) and odds ratios (OR). \\

\textbf{Findings:} Progressive PPFE independently associated with mortality in both LCS cohorts (NLST: Hazard Ratio (HR)=1.25, 95\% Confidence Interval (CI): 1.01--1.56, $\text{p=0.042}$; SUMMIT: HR=3.14, 95\% CI: 1.66--5.97, $\text{p}\raisebox{0.15ex}{$<$}0.001$). Within SUMMIT, progressive PPFE was strongly associated with higher respiratory admissions (IRR=2.79, $\text{p}\raisebox{0.15ex}{$<$}0.001$), increased antibiotic and steroid use (IRR=1.55, p=0.010), and showed a trend towards higher modified medical research council scores (OR=1.40, p=0.055). \\

\textbf{Interpretation:} Radiologic PPFE progression independently associates with mortality across two large LCS cohorts, and associates with adverse clinical outcomes. Quantitative assessment of PPFE progression may provide a clinically relevant imaging biomarker to identify individuals at increased risk of respiratory morbidity within LCS programmes.

\end{abstract}

\end{frontmatter}

\section{Introduction}
Pleuroparenchymal fibroelastosis (PPFE) is a rare interstitial pneumonia that is characterised on histopathology by the presence of pleural and intraparenchymal fibroelastosis. PPFE can occur as an idiopathic entity or may occur in the presence of fibrosing interstitial lung diseases (ILD), such as idiopathic pulmonary fibrosis (IPF)~\citep{jacob2018functional,gudmundsson2021pleuroparenchymal,gudmundsson2023delineating} systemic sclerosis~\citep{bonifazi2020pleuroparenchymal}, or fibrosing hypersensitivity pneumonitis (HP)~\citep{jacob2018functional}. Radiologically, PPFE is visible on computed tomography (CT) as dense triangular or band-like subpleural opacities and predominately affect the upper lobes ~\citep{amitani1992idiopathic,frankel2004idiopathic,watanabe2013pleuroparenchymal,raghu2018diagnosis}. 

Idiopathic PPFE has been found to associate with increased mortality when occurring in isolation on CT imaging~\citep{shioya2018poorer,tanizawa2018clinical,fukada2022idiopathic}. However, in recent years, its independent impact on survival has been reported in patients with fibrosing ILDs~\citep{gudmundsson2023delineating,jacob2018functional,gudmundsson2021pleuroparenchymal,bonifazi2020pleuroparenchymal,fujisawa2021radiological}. Quantitative analysis of upper lobe PPFE on baseline CT imaging, has been shown to have greater sensitivity for the detection of prognostically important PPFE when compared to visual PPFE estimation~\citep{gudmundsson2021pleuroparenchymal,gudmundsson2023delineating}. Furthermore, longitudinal CT quantification of PPFE progression in patients with IPF and HP has shown strong associations with mortality independent of change in forced vital capacity (FVC) or change in ILD extent scored visually on CT~\citep{gudmundsson2023delineating}. These findings suggest that upper-lobe PPFE progression may develop untethered to co-existing usual interstitial pneumonia (UIP) type fibrosis that predominantly evolves at the lung bases.

The National Lung Screening Trial (NLST)~\citep{national2011national, national2011reduced} based in the United States of America and the SUMMIT~\citep{dickson2023uptake} study based in London are lung cancer screening (LCS) studies that enrolled people (current or former smokers) deemed to be at high risk of developing lung cancer. Participants underwent low-dose CT (LDCT) screening as a means to accelerate early lung cancer detection and thereby reduce lung cancer-related mortality. LCS populations such as NLST~\citep{national2011national, national2011reduced} and SUMMIT~\citep{dickson2023uptake} provide a unique opportunity to study the prevalence and progression of rare chronic lung diseases such as PPFE in a population enriched for prior/current heavy smokers. Given that the National Health Service in the UK estimates that 1 million participants will undergo annual CT screening by 2028 as part of LCS programs, understanding the impact of PPFE on survival in LCS populations has developed increased clinical relevance. Accordingly, in this study, we aimed to investigate the prognostic impact of automated quantification of radiologic PPFE progression in longitudinal LDCT scans from NLST~\citep{national2011national, national2011reduced} and SUMMIT~\citep{dickson2023uptake}.

\section{Methods}
\subsection*{Study population}
\begin{figure*}[ht]
\centering
\includegraphics[width=\linewidth]{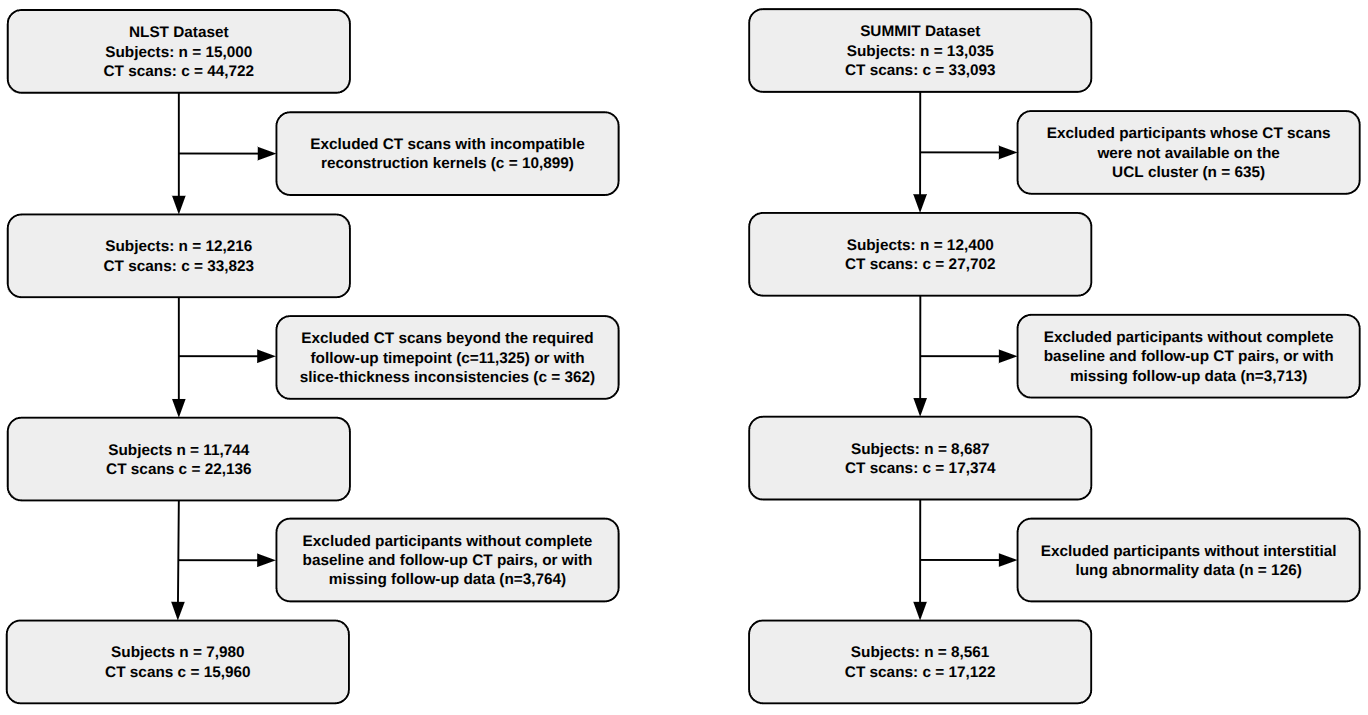}
\caption{Flowchart illustrating the selection of participants from the NLST and SUMMIT datasets. From the NLST cohort (n = 15,000; 44,722 CT scans), scans reconstructed using both lung and soft-tissue kernels were retained, followed by the selection of participants with baseline and one-year follow-up CTs and available survival data, yielding a final cohort of 7,980 participants (15,960 CT scans). From the SUMMIT cohort (n = 13,035; 33,093 CT scans), participants with baseline and two-year follow-up CTs and available follow-up information were included. After excluding individuals without ILA data, the final SUMMIT cohort comprised 8,561 participants (17,122 CT scans).}
\label{flowchart_change}
\end{figure*}

The NLST~\citep{national2011national,national2011reduced} was a large, randomized, multicentre lung cancer screening trial conducted in the United States that ran from August 2002 to September 2007, in which 26,722 participants underwent three annual LDCT screenings. A subpopulation of 15,000 participants (44,772 scans) from the NLST was available for evaluation in this study. This subset included individuals with multiple timepoint volumetric LDCT scans and up to ten years of follow-up survival data (collected in 2015). All NLST LDCT scans were acquired using multidetector CT scanners with varying numbers of detector channels. To ensure image quality and consistency, specific selection criteria were applied in this study (Figure~\ref{flowchart_change}). Clinical data collected included age, sex, smoking history (pack-years), and height. Anonymized data retrospectively analysed in this study were obtained with informed consent from all participants enrolled in the NLST~\citep{national2011national,national2011reduced}. The analysis of NLST data in this study was approved by the UCL Research Ethics Committee (15401/002), and all research was conducted in accordance with the principles of the Declaration of Helsinki.

The SUMMIT~\citep{dickson2023uptake} study was a North London–based LCS programme designed to evaluate the implementation of LDCT screening for the early detection of lung cancer in a high-risk population, and to validate a multi-cancer early detection blood test (ClinicalTrials.gov identifier:NCT03934866).
The SUMMIT screening programme enrolled and analysed 13,035 current and former smokers aged 55–77 years, using multiple LDCT timepoints. The first LDCT screening was performed after the baseline Lung Health Check (LHC) visit, unless participants reported symptoms of an active lower respiratory tract infection (in which case scanning was deferred for six weeks) or requested a later appointment. Two years after baseline, all participants were invited for a follow-up LDCT scan. Demographic, smoking, and medical history data were collected at the LHC appointments, including self-reported age, sex, smoking history (pack-years), height, spirometry measurements, modified Medical Research Council (mMRC) dyspnoea score, respiratory hospitalisations, and steroid and antibiotic use for chest symptoms. Respiratory hospitalisations were defined as any hospital admission with a primary or secondary diagnosis coded under the International Classification of Diseases, 10th Revision (ICD-10) “Diseases of the Respiratory System,” identified using Hospital Episode Statistics from NHS England at year 1 and year 2. Age and sex were obtained from primary care records \citep{cheng2026fibrotic}. Visual assessment of interstitial lung abnormalities (ILA) was performed in the SUMMIT cohort by expert thoracic radiologists using a predefined classification scheme \citep{cheng2026fibrotic}; full details are provided in the Supplementary Appendix. Written informed consent was obtained from all participants in the SUMMIT study following determination of study eligibility. Ethical approval was granted by an NHS Research Ethics Committee (17/LO/2004) and the NHS Health Research Authority’s Confidentiality Advisory Group (18/CAG/0054).

\subsection*{PPFE Quantification} 
Automated PPFE quantification was performed using a deep learning–based approach to segment upper-lung PPFE on LDCT scans; detailed implementation and training procedures are provided in Supplementary Appendix. Radiological PPFE lesions were automatically segmented on baseline and follow-up LDCT scans. The quantitative volume of PPFE for each scan was calculated on a continuous scale as the number of voxels segmented as PPFE multiplied by the dimensions of the voxels to obtain a total volume in cubic centimeters (cm$^3$). Longitudinal change in PPFE ($\Delta$PPFE) was defined as the difference between follow-up and baseline PPFE volume divided by the time interval between scans (cm$^3$/year). To characterise progression, we derived a distribution-based threshold using half of the standard deviation of baseline PPFE in the NLST cohort, consistent with established biomarker methodology and representing a moderate effect size \citep{gudmundsson2023delineating,norman2003interpretation,swigris20106,humphries2018quantitative}. The numerical value corresponding to this threshold was calculated within NLST and subsequently applied to the SUMMIT cohort. Progressive PPFE was analysed as a binary progression phenotype defined by this derived threshold in all inferential analyses.

\subsection*{Statistical Analysis}
Baseline demographic and clinical characteristics were summarised for each cohort and compared between participants with progressive PPFE and those without progression. Continuous variables were assessed for group differences using Welch’s independent two-sample t-tests, and categorical variables were compared using Pearson’s chi-squared tests of independence. Overall survival was visualised using Kaplan–Meier curves stratified by PPFE progression status, and survival distributions were compared using the log-rank test.

Survival associations with all-cause mortality were then evaluated using Cox proportional hazards (Cox-PH) regression models in the SUMMIT and NLST cohorts. Univariable Cox models were first fitted for each covariate to estimate hazard ratios (HRs) and concordance indices (C-index). Multivariable Cox models were subsequently constructed, adjusting for age, sex, smoking history (pack-years), height (cm), baseline PPFE on the first scan, and the interaction between baseline PPFE and progressive PPFE. In SUMMIT, FVC \% predicted and ILA status were additionally included. Model discrimination was assessed using the C-index, and analyses were performed separately in each cohort.

Finally, within the SUMMIT cohort, regression analyses were performed to assess associations between progressive PPFE and key clinical outcomes. Respiratory hospital admissions were modelled using negative binomial generalised linear models to account for overdispersed count data. Dyspnoea severity was assessed using ordinal logistic regression with two-year mMRC score as the dependent variable. Antibiotic and/or steroid use for chest symptoms was analysed using negative binomial regression, with annual and cumulative prescription counts as dependent variables. All secondary models included progressive PPFE as the main predictor and were adjusted for age, sex, smoking history (pack-years), FVC \% predicted, ILA, and baseline PPFE. Results were reported as incidence rate ratios (IRRs) or odds ratios (ORs) with 95\% confidence intervals. All continuous variables were modelled per one-unit increase in their native measurement scale. All statistical analyses were performed in Python (version 3.11) using the lifelines, statsmodels, and scikit-survival libraries.

To evaluate potential selection bias arising from excluding participants without longitudinal CT imaging, we conducted a secondary analysis within the SUMMIT cohort, restricted to 2,580 participants with only baseline CT. In this subgroup, associations between baseline PPFE and clinical outcomes were assessed using the same modelling framework as the main analysis (Supplementary Table \ref{tab_demographics_baseline} and Figure \ref{flowchart_baseline}).

\section{Results}
\subsection*{Comparison of study populations}
A total of 7,980 participants from the NLST cohort and 8,561 participants from the SUMMIT study with appropriate longitudinal CTs were included in the cardinal analyses (Table~\ref{tab_demographics_change}). In the NLST cohort, 431(5.4\%) participants were designated as progressive PPFE. In the SUMMIT cohort, 124 participants (1.5\%) showed progressive PPFE. Participants in NLST were on average 4 years younger than those in SUMMIT, but had an 8-year longer pack-year smoking history. Whilst sex distributions were similar across cohorts, SUMMIT participants had more PPFE at baseline and follow-up, but the rate of change in subjects with progressive PPFE was higher in NLST (Table~\ref{tab_demographics_change}). In NLST, baseline and follow-up PPFE values were 0.70±1.47 and 1.38±1.62, respectively, with a $\Delta$PPFE of 0.69±0.37 in progressive participants. In SUMMIT, mean PPFE increased from 2.39±2.51 at baseline to 3.83±2.54 at follow-up, with an annualised change of 0.65±0.24 in progressive cases.

\begin{figure}
\centering
\includegraphics[width=\linewidth]{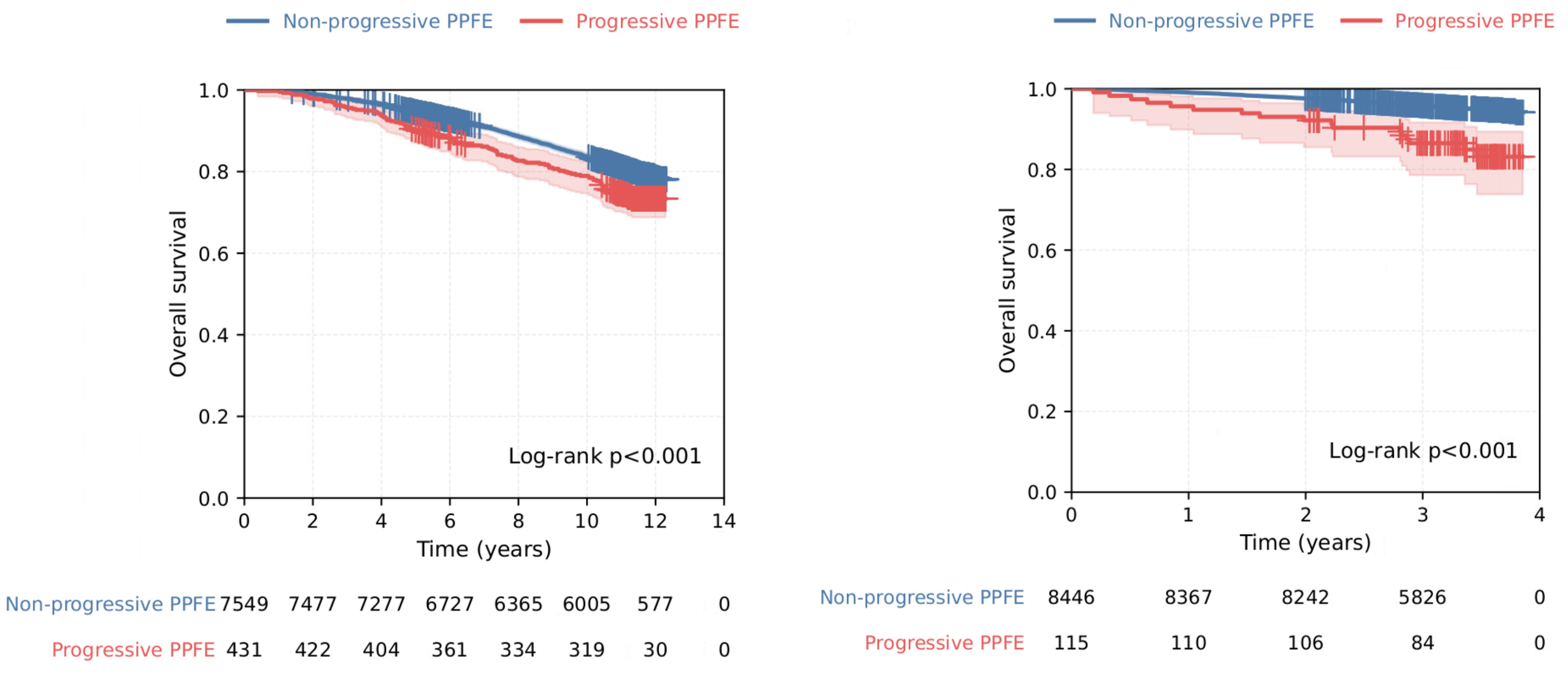}
\caption{Kaplan–Meier survival by longitudinal PPFE progression. NLST (ten-year follow-up) and SUMMIT (two-year follow-up) cohorts comparing overall survival between participants with progressive and non-progressive PPFE change. Shaded areas indicate 95\% confidence intervals and tick marks denote censoring. Numbers at risk are displayed below each plot. Participants with progressive PPFE exhibited significantly reduced survival in both cohorts (log-rank \text{p}\raisebox{0.15ex}{$<$}0.001 for both NLST and SUMMIT), consistent with multivariable Cox models (Table \ref{tab_multi_cox_two_cohorts}) in which progressive PPFE remained independently associated with mortality.}
\label{Kaplan_Meier_Delta_PPFE}
\end{figure}

\begin{table*}[htbp]
\centering
\caption{Patient demographics, and clinical variables including age, sex, height, smoking history (pack-years), FVC, visual scores of ILA, and computer-based scores of PPFE in NLST and SUMMIT cohorts. PPFE (baseline): upper-lung pleuroparenchymal fibroelastosis quantified on the first scan, PPFE (follow-up): upper-lung pleuroparenchymal fibroelastosis quantified on the follow-up scan, $\Delta$PPFE: annualised change in computerised upper-lung PPFE between scans, ILA: visual scores of interstitial lung abnormality, FVC: forced vital capacity, Progressive: $\Delta$PPFE $\geq$ 0.41 cm$^3$/year. PPFE (baseline), PPFE (follow-up), $\Delta$PPFE, Height, and FVC are described as Mean±SD. Age and smoking are described as medians (range). Sex is described as Female/Male \%. ILA is described as number/\%.}
\begingroup
\scriptsize
\setlength{\tabcolsep}{4pt}
\renewcommand{\arraystretch}{1.2}
\begin{tabular}{lcccccc}
\toprule
\textbf{Variable} &
\multicolumn{3}{c}{\textbf{NLST}} &
\multicolumn{3}{c}{\textbf{SUMMIT}} \\
\cmidrule(lr){2-4} \cmidrule(lr){5-7}
& \textbf{Non-Progressive} & \textbf{Progressive*} & \textbf{p-value}
& \textbf{Non-Progressive} & \textbf{Progressive*} & \textbf{p-value} \\
\midrule
Num Patients, (\%) 
& 7549 (94.6) & 431 (5.4) & --
& 8437 (98.5) & 124 (1.5) & -- \\
Age y, median (range)
& 61 (55--74) & 63 (54--74) & $\raisebox{0.15ex}{$<$}0.001$
& 65 (55--80) & 66 (55--77) & 0.104 \\
Sex (F/M\%)
& 42/58 & 36/64 & 0.020
& 42/58 & 31/69 & 0.021 \\
Smoking py, median (range)
& 48 (30--272) & 51 (21--162) & 0.007
& 40 (2.5--400) & 43 (16.5--310) & 0.226 \\
FVC \%pred, mean$\pm$SD
& -- & -- & --
& 88.12$\pm$16.51 & 84.92$\pm$18.13 & 0.053 \\
Height cm, mean$\pm$SD
& 172.35$\pm$9.95 & 174.10$\pm$10.19 & $\raisebox{0.15ex}{$<$}0.001$
& 168.89$\pm$9.77 & 171.71$\pm$10.50 & 0.003 \\
ILA, n(\%)
& -- & -- & --
& 1316 (15.6) & 42 (33.8) & $\raisebox{0.15ex}{$<$}0.001$ \\
PPFE (baseline), mean$\pm$SD
& 0.27$\pm$0.76 & 0.70$\pm$1.47 & $\raisebox{0.15ex}{$<$}0.001$
& 0.43$\pm$1.01 & 2.39$\pm$2.51 & $\raisebox{0.15ex}{$<$}0.001$ \\
PPFE (follow-up), mean$\pm$SD
& 0.29$\pm$0.68 & 1.38$\pm$1.62 & $\raisebox{0.15ex}{$<$}0.001$
& 0.51$\pm$0.97 & 3.83$\pm$2.54 & $\raisebox{0.15ex}{$<$}0.001$ \\
$\Delta$PPFE, mean$\pm$SD
& -- & 0.69$\pm$0.37 & --
& -- & 0.65$\pm$0.24 & -- \\
\bottomrule
\end{tabular}
\par\smallskip
\begin{flushleft}
\scriptsize Progressive* = $\Delta$PPFE $\geq$ 0.41 cm$^3$/year.
\end{flushleft}
\endgroup
\label{tab_demographics_change}
\end{table*}

\begin{table*}[htbp]
\centering
\caption{Association of progressive PPFE with mortality in univariable Cox proportional hazards models in NLST and SUMMIT cohorts. Univariable models fitted separately for each cohort. Models adjusted for a single covariate at a time to assess its univariate association with mortality. C-index values represent model discrimination for each variable. HRs in bold indicate p < 0.05. Baseline PPFE: upper-lung pleuroparenchymal fibroelastosis quantified on the first scan, Progressive PPFE: $\Delta$PPFE $\geq$ 0.41 cm$^3$/year, ILA: visual scores of interstitial lung abnormality, FVC: forced vital capacity.}
\begingroup
\scriptsize
\setlength{\tabcolsep}{6pt}
\renewcommand{\arraystretch}{1.12}
\begin{tabular}{lcccc}
\toprule
\textbf{Variable} & \textbf{Hazard ratio} & \textbf{95\% confidence interval} & \textbf{p-value} & \textbf{C-index} \\
\midrule
\multicolumn{5}{l}{\textbf{NLST}} \\
Age (y)                              & \textbf{1.10} & 1.09--1.11 & $\raisebox{0.15ex}{$<$}0.001$ & 0.63 \\
Sex (male)                        & \textbf{1.49} & 1.34--1.66 & $\raisebox{0.15ex}{$<$}0.001$ & 0.54 \\
Smoking (py)                         & \textbf{1.01} & 1.01--1.01 & $\raisebox{0.15ex}{$<$}0.001$ & 0.60 \\
Height (cm)                          & \textbf{1.01} & 1.00--1.01 & $\raisebox{0.15ex}{$<$}0.001$  & 0.54 \\
Baseline PPFE                        & \textbf{1.06} & 1.01--1.12 & 0.015  & 0.52 \\
Progressive PPFE                        & \textbf{1.37} & 1.13--1.67 & $\raisebox{0.15ex}{$<$}0.001$  & 0.51 \\
\midrule
\multicolumn{5}{l}{\textbf{SUMMIT}} \\
Age (y)                              & \textbf{1.09} & 1.07--1.11 & $\raisebox{0.15ex}{$<$}0.001$ & 0.65 \\
Sex (male)                        & \textbf{1.51} & 1.22--1.86 & $\raisebox{0.15ex}{$<$}0.001$  & 0.54 \\
Smoking (py)                         & \textbf{1.00} & 1.00--1.01 & $\raisebox{0.15ex}{$<$}0.001$  & 0.56 \\
ILA (present vs absent)              & \textbf{1.81} & 1.45--2.27 & $\raisebox{0.15ex}{$<$}0.001$  & 0.55 \\
Height (cm)                          & 1.00          & 0.99--1.01 & 0.388  & 0.51 \\
FVC \% predicted                     & \textbf{0.98} & 0.97--0.98 & $\raisebox{0.15ex}{$<$}0.001$ & 0.59 \\
Baseline PPFE                        & \textbf{1.12} & 1.04--1.20 & $\raisebox{0.15ex}{$<$}0.001$  & 0.54 \\
Progressive PPFE                         & \textbf{3.25} & 2.02--5.21 & $\raisebox{0.15ex}{$<$}0.001$  & 0.52 \\
\bottomrule
\end{tabular}
\par\smallskip
\caption*{}
\endgroup
\label{tab_univariate_cox_two_cohorts}
\end{table*}

\subsection*{Comparisons of progressive PPFE}
The distribution-based threshold used to define progressive PPFE was derived as half the standard derivation of baseline PPFE measured on the NLST cohort (n=7980; Baseline PPFE=0.35±0.81; Table \ref{tab_cohort_summary_all} Supplementary Appendix). Progressive PPFE was therefore defined as $\Delta$PPFE$\geq$0.41 cm$^3$/year. Group-wise comparisons between participants with progressive and non-progressive PPFE were performed separately for each cohort (Table~\ref{tab_demographics_change}). Across both cohorts, participants with progressive PPFE were older (NLST:$\text{p}\raisebox{0.15ex}{$<$}0.001$; SUMMIT:p=0.104) and had greater smoking exposure (NLST:p=0.007; SUMMIT:p=0.226) compared with those without progression. Participants with progressive PPFE were also taller (NLST: $\text{p}\raisebox{0.15ex}{$<$}0.001$; SUMMIT:p=0.003). Progressive PPFE cases demonstrated a substantially greater PPFE burden at baseline in both NLST and SUMMIT. In the SUMMIT cohort, participants with progressive PPFE had lower baseline FVC \% predicted compared with non-progressive participants (84.92±18.13\% vs 88.12±16.51\%, p=0.053). The prevalence of ILA in SUMMIT was markedly higher among participants with progressive PPFE (42/124, 33.8\%) compared with those without progression (1316/8437, 15.6\%; $\text{p}\raisebox{0.15ex}{$<$}0.001$).

\subsection*{Mortality Analysis}
In univariable Cox proportional hazards models (Table~\ref{tab_univariate_cox_two_cohorts}), older age and male sex were strongly associated with higher mortality in both cohorts. Progressive PPFE showed a significant association with mortality (HR=1.37, 95\% CI: 1.13–1.67, $\text{p}\raisebox{0.15ex}{$<$}0.001$) in NLST, with a stronger association  observed in the SUMMIT cohort (HR=3.25, 95\% CI: 2.02–5.21, $\text{p}\raisebox{0.15ex}{$<$}0.001$). Baseline PPFE also demonstrated a weaker but significant relationship with mortality in both datasets.

Kaplan--Meier survival curves (Figure~\ref{Kaplan_Meier_Delta_PPFE}) demonstrated that in both cohorts, participants with progressive PPFE showed reduced overall survival compared to those without progressive PPFE. In the SUMMIT cohort, separation between curves was evident within the first two years and persisted throughout follow-up (log-rank:$\text{p}\raisebox{0.15ex}{$<$}0.001$), while in NLST a similar but less pronounced divergence was observed over ten years of follow-up (log-rank:$\text{p}\raisebox{0.15ex}{$<$}0.001$). 

In multivariable Cox proportional hazards models adjusted for age, sex, smoking history (pack-years), height, baseline PPFE, and the interaction between baseline PPFE and progressive PPFE (Table~\ref{tab_multi_cox_two_cohorts}), progressive PPFE remained independently associated with mortality in both cohorts (NLST:HR=1.25, 95\% CI: 1.01–1.56, p=0.042); SUMMIT:HR=3.14, 95\% CI: 1.66–5.97, $\text{p}\raisebox{0.15ex}{$<$}0.001$). In both models, age and male sex were consistent predictors of increased mortality, while smoking pack-years showed smaller but significant effects. Baseline PPFE and height were not significantly associated with mortality after adjusting for other covariates. When the model was further adjusted for visual ILA scores and FVC \% predicted in the SUMMIT cohort (Table~\ref{tab_multivariable_cox_summit}), progressive PPFE remained independently associated with mortality (HR=2.55, 95\% CI: 1.34–4.85, p=0.004).

\begin{table*}[htbp]
\centering
\caption{Association of progressive PPFE with mortality in multivariable Cox proportional hazards models in NLST and SUMMIT cohorts. Models adjusted for age, sex, smoking (pack-years), height, baseline PPFE, and the interaction between baseline PPFE and progressive PPFE. C-index shown per cohort for the multivariable model. HRs in bold indicate p < 0.05. Baseline PPFE: upper-lung pleuroparenchymal fibroelastosis quantified on the first scan, Progressive PPFE: $\Delta$PPFE $\geq$ 0.41 cm$^3$/year.}
\begingroup
\scriptsize
\setlength{\tabcolsep}{6pt}
\renewcommand{\arraystretch}{1.15}
\begin{tabular}{lcccc}
\toprule
\textbf{Variable} & \textbf{Hazard ratio} & \textbf{95\% confidence interval} & \textbf{p-value} & \textbf{Model C-index} \\
\midrule
\multicolumn{5}{l}{\textbf{NLST}} \\
Age (y)                         & \textbf{1.10} & 1.09--1.11 & $\raisebox{0.15ex}{$<$}0.001$ & 0.66 \\
Sex (male)                   & \textbf{1.30} & 1.11--1.51 & $\raisebox{0.15ex}{$<$}0.001$  &  \\
Smoking (py)                    & \textbf{1.01} & 1.01--1.01 & $\raisebox{0.15ex}{$<$}0.001$ &  \\
Height (cm)                     & 1.00          & 0.99--1.01 & 0.925  &  \\
Baseline PPFE                   & 1.02          & 0.96--1.08 & 0.552  &  \\
Progressive PPFE                    & \textbf{1.25} & 1.01--1.56 & 0.042  &  \\
Progressive PPFE $\times$ Baseline PPFE (interaction) & 0.92 & 0.79--1.08 & 0.331 &  \\
\midrule
\multicolumn{5}{l}{\textbf{SUMMIT}} \\
Age (y)                         & \textbf{1.08} & 1.07--1.10 & $\raisebox{0.15ex}{$<$}0.001$ & 0.67 \\
Sex (male)                   & \textbf{1.57} & 1.21--2.05 & $\raisebox{0.15ex}{$<$}0.001$  &  \\
Smoking (py)                    & \textbf{1.00} & 1.00--1.01 & 0.006  &  \\
Height (cm)                     & 0.99          & 0.98--1.01 & 0.327                &  \\
Baseline PPFE                   & 1.06          & 0.98--1.15 & 0.095              &  \\
Progressive PPFE                    & \textbf{3.14} & 1.66--5.97 & $\raisebox{0.15ex}{$<$}0.001$  &  \\
Progressive PPFE $\times$ Baseline PPFE (interaction) & 0.90 & 0.72--1.12 & 0.347 &  \\
\bottomrule
\end{tabular}
\par\smallskip
\caption*{}
\endgroup
\label{tab_multi_cox_two_cohorts}
\end{table*}

\subsection*{Clinical Outcomes in SUMMIT}
We examined the association between progressive PPFE and various clinical outcomes in SUMMIT, including respiratory admissions, mMRC score, and steroid/antibiotic medication use. All models were adjusted for age, sex, smoking history (pack-years), FVC \% predicted, visually assessed ILA and baseline PPFE.

In the negative binomial GLM model (Table~\ref{tab_neg_binomial_glm_summit-respiratory}), progressive PPFE was significantly associated with higher rates of respiratory admissions (IRR=2.79, 95\% CI: 1.69–4.60, $\text{p}\raisebox{0.15ex}{$<$}0.001$). Greater baseline PPFE extent was also independently associated with a higher incidence of respiratory admissions (IRR=1.13, 95\% CI: 1.06–1.22, $\text{p}\raisebox{0.15ex}{$<$}0.001$). Older age and smoking pack-years were also associated with increased admission rates, whereas male sex and higher FVC\% predicted were inversely associated with admissions.

Ordinal logistic regression (Table~\ref{tab_ordinal_logistic_summit- mMRC}) demonstrated that both progressive and baseline PPFE extent were associated with higher odds of more severe mMRC symptom scores. Baseline PPFE showed a consistent association with symptom severity (OR=0.95, 95\% CI: 0.91–0.99, p=0.007), while the estimate for progressive PPFE was less precise, with wider confidence intervals (OR=1.40, 95\% CI: 0.99–1.97, p=0.055), indicating greater uncertainty around the magnitude of its association with respiratory symptoms. Older age, higher smoking exposure, and the presence of ILA were associated with more severe symptoms, whereas male sex and higher FVC\% predicted were associated with lower symptom burden.

In the negative binomial model for steroid or antibiotic use (Table~\ref{tab_neg_binomial_steroid_summit}), progressive PPFE remained a significant predictor of higher medication use (IRR=1.55, 95\% CI: 1.10–2.19, p=0.011). Baseline PPFE was also positively associated with increased steroid or antibiotic use (IRR=1.06, 95\% CI: 1.02–1.10, p=0.006). Again, male sex and higher FVC\% predicted were inversely associated with inflammatory medication use.
Across all models, replacing FVC\% predicted with FEV$_1$\% predicted yielded comparable estimates and did not materially alter the results.

In analyses of Major Adverse Cardiovascular Events (MACE5), defined as a composite of non-fatal myocardial infarction, stroke, coronary revascularisation, heart failure hospitalisation, or cardiovascular death in SUMMIT, neither baseline nor progressive PPFE showed a significant association with MACE5 after adjustment for covariates (all $\text{p}\raisebox{0.15ex}{$>$}0.05$). These findings suggest that in this screening population, radiologic PPFE burden and its progression, primarily relate to adverse respiratory rather than cardiovascular disease outcomes.

In a secondary analysis undertaken to assess potential bias from excluding participants without longitudinal imaging, 2,580 SUMMIT participants with baseline CT alone were evaluated. In this subgroup, greater baseline PPFE remained independently associated with higher respiratory admission rates (IRR=1.24, 95\% CI: 1.17–1.31, p<0.001), but was not associated with mMRC dyspnoea score (OR=1.01, p=0.701) or steroid/antibiotic use (IRR=0.96, p=0.436) (Tables \ref{tab_neg_binomial_respiratory_summit} - \ref{tab_neg_binomial_steroid_y1_summit}). These findings suggest that baseline PPFE captures some respiratory risk even in the absence of follow-up imaging, but that longitudinal PPFE change provides stronger associations with symptomatic and treatment-related outcomes.

\begin{table*}[htbp]
\centering
\caption{Multivariable Cox proportional hazards model assessing progressive PPFE and mortality in the SUMMIT cohort. Model adjusted for age, sex, smoking (pack-years), height, FVC \% predicted, ILA, baseline PPFE, and the interaction between baseline PPFE and progressive PPFE. C-index shown for the multivariable model. HRs in bold indicate p < 0.05. Baseline PPFE: upper-lung pleuroparenchymal fibroelastosis quantified on the first scan, Progressive PPFE: $\Delta$PPFE $\geq$ 0.41 cm$^3$/year, ILA: visual scores of interstitial lung abnormality, FVC: forced vital capacity.}
\begingroup
\scriptsize
\setlength{\tabcolsep}{6pt}
\renewcommand{\arraystretch}{1.12}
\begin{tabular}{lcccc}
\toprule
\textbf{Variable} & \textbf{Hazard ratio} & \textbf{95\% confidence interval} & \textbf{p-value} & \textbf{C-index} \\
\midrule
\multicolumn{5}{l}{\textbf{SUMMIT}} \\
Age (y)                             & \textbf{1.08} & 1.06--1.10 & $\raisebox{0.15ex}{$<$}0.001$ & 0.70 \\
Sex (male)                       & \textbf{1.45} & 1.11--1.89 & 0.005  &      \\
Smoking (py)                        & 1.00          & 1.00--1.01 & 0.051  &      \\
ILA (present vs absent)             & \textbf{1.42} & 1.13--1.79 & 0.002  &      \\
Height (cm)                         & 1.00          & 0.98--1.01 & 0.431  &      \\
FVC \% predicted                    & \textbf{0.98} & 0.98--0.99 & $\raisebox{0.15ex}{$<$}0.001$ &      \\
Baseline PPFE                       & 1.07          & 0.99--1.16 & 0.077  &      \\
Progressive PPFE                        & \textbf{2.55} & 1.34--4.85 & 0.004  &      \\
Progressive PPFE $\times$ Baseline PPFE (interaction) & 0.92 & 0.74--1.14 & 0.441 & \\
\bottomrule
\end{tabular}
\par\smallskip
\caption*{}
\endgroup
\label{tab_multivariable_cox_summit}
\end{table*}

\begin{table*}[htbp]
\centering
\caption{Negative Binomial GLM model for respiratory admissions in the SUMMIT cohort. Model adjusted for age, sex, smoking (pack-years), FVC \% predicted, ILA, and baseline PPFE. IRRs in bold indicate p < 0.05. Baseline PPFE: upper-lung pleuroparenchymal fibroelastosis quantified on the first scan, Progressive PPFE: $\Delta$PPFE $\geq$ 0.41 cm$^3$/year, ILA: visual scores of interstitial lung abnormality, FVC: forced vital capacity.}
\begingroup
\scriptsize
\setlength{\tabcolsep}{6pt}
\renewcommand{\arraystretch}{1.12}
\begin{tabular}{lccc}
\toprule
\textbf{Variable} & \textbf{Incidence Rate Ratio} & \textbf{95\% confidence interval} & \textbf{p-value} \\
\midrule
Age (y)                 & \textbf{1.051} & 1.033--1.070 & $\raisebox{0.15ex}{$<$}0.001$ \\
Sex (male)           & \textbf{0.696} & 0.563--0.860 & $\raisebox{0.15ex}{$<$}0.001$ \\
Smoking (py)            & \textbf{1.005} & 1.002--1.009 & $\raisebox{0.15ex}{$<$}0.001$ \\
FVC \% predicted        & \textbf{0.967} & 0.962--0.973 & $\raisebox{0.15ex}{$<$}0.001$ \\
ILA (present vs absent)      & 1.275          & 0.979--1.661 & 0.070 \\
Baseline PPFE      & \textbf{1.135} & 1.059--1.217 & $\raisebox{0.15ex}{$<$}0.001$ \\
Progressive PPFE       & \textbf{2.786} & 1.688--4.596 & $\raisebox{0.15ex}{$<$}0.001$ \\
\bottomrule
\end{tabular}
\par\smallskip
\caption*{}
\endgroup
\label{tab_neg_binomial_glm_summit-respiratory}
\end{table*}

\begin{table*}[htbp]
\centering
\caption{Ordinal logistic regression model for mMRC score in the SUMMIT cohort. Model adjusted for age, sex, smoking (pack-years), FVC \% predicted, ILA, and baseline PPFE. Odds in bold indicate p < 0.05. Baseline PPFE: upper-lung pleuroparenchymal fibroelastosis quantified on the first scan, Progressive PPFE: $\Delta$PPFE $\geq$ 0.41 cm$^3$/year, ILA: visual scores of interstitial lung abnormality, FVC: forced vital capacity.}
\begingroup
\scriptsize
\setlength{\tabcolsep}{6pt}
\renewcommand{\arraystretch}{1.12}
\begin{tabular}{lccc}
\toprule
\textbf{Variable} & \textbf{Odds Ratio} & \textbf{95\% confidence interval} & \textbf{p-value} \\
\midrule
Age (y)                & \textbf{1.021} & 1.014--1.028 & $\raisebox{0.15ex}{$<$}0.001$ \\
Sex (male)                & \textbf{0.554} & 0.509--0.601 & $\raisebox{0.15ex}{$<$}0.001$ \\
Smoking (pack-years)    & \textbf{1.012} & 1.010--1.013 & $\raisebox{0.15ex}{$<$}0.001$ \\
FVC \% predicted        & \textbf{0.972} & 0.970--0.975 & $\raisebox{0.15ex}{$<$}0.001$ \\
ILA (present vs absent)    & \textbf{1.126} & 1.134--1.417 & $\raisebox{0.15ex}{$<$}0.001$ \\
Baseline PPFE          & \textbf{0.949} & 0.913--0.986 & 0.007 \\
Progressive PPFE           & 1.400 & 0.993--1.974 & 0.055 \\
\bottomrule
\end{tabular}
\par\smallskip
\caption*{}
\endgroup
\label{tab_ordinal_logistic_summit- mMRC}
\end{table*}

\begin{table*}[htbp]
\centering
\caption{Negative Binomial GLM model for steroid/antibiotic inflammatory medication use in the SUMMIT cohort. Model adjusted for age, sex, smoking (pack-years), FVC \% predicted, ILA, and baseline PPFE. IRRs in bold indicate p < 0.05. Baseline PPFE: upper-lung pleuroparenchymal fibroelastosis quantified on the first scan, Progressive PPFE: $\Delta$PPFE $\geq$ 0.41 cm$^3$/year, ILA: visual scores of interstitial lung abnormality, FVC: forced vital capacity.}
\begingroup
\scriptsize
\setlength{\tabcolsep}{6pt}
\renewcommand{\arraystretch}{1.12}
\begin{tabular}{lccc}
\toprule
\textbf{Variable} & \textbf{Incidence Rate Ratio} & \textbf{95\% confidence interval} & \textbf{p-value} \\
\midrule
Age (y)                & 0.994 & 0.986--1.002 & 0.201 \\
Sex (male)               & \textbf{0.540} & 0.490--0.595 & $\raisebox{0.15ex}{$<$}0.001$ \\
Smoking (pack-years)    & \textbf{1.007} & 1.005--1.009 & $\raisebox{0.15ex}{$<$}0.001$ \\
FVC \% predicted        & \textbf{0.982} & 0.979--0.985 & $\raisebox{0.15ex}{$<$}0.001$ \\
ILA (present vs absent)      & 1.009 & 0.881--1.155 & 0.891 \\
Baseline PPFE         & \textbf{1.059} & 1.016--1.104 & 0.006 \\
Progressive PPFE          & \textbf{1.555} & 1.105--2.190 & 0.011 \\
\bottomrule
\end{tabular}
\par\smallskip
\caption*{}
\endgroup
\label{tab_neg_binomial_steroid_summit}
\end{table*}

\section{Discussion}
Our study, demonstrates across two independent lung cancer screening datasets, that radiologic progression of PPFE is independently associated with increased all-cause mortality. This association was consistent across two demographically and chronologically distinct screening populations, SUMMIT and NLST, despite differences in cohort design, image acquisition protocols and scanner model generations, reconstruction techniques, scan  intervals and follow-up duration. Importantly, progressive PPFE remained a strong predictor of mortality even after adjustment for demographic factors, smoking history (pack-years), height, baseline PPFE, and, in SUMMIT, for both ILA and FVC. These results support the concept that radiologic PPFE progression captures a biological process associated with respiratory morbidity that is distinct from and additive to established risk factors measured in lung cancer screening populations.

Our findings extend prior work in established fibrosing ILDs such as IPF and hypersensitivity pneumonitis,~\citep{jacob2018functional, gudmundsson2021pleuroparenchymal, gudmundsson2023delineating, bonifazi2020pleuroparenchymal} and show that PPFE progression is prognostically important in asymptomatic or minimally symptomatic individuals undergoing lung cancer screening. We also observed that baseline PPFE demonstrated weaker and less consistent prognostic value compared with progressive PPFE. This interpretation was supported by a subanalysis of SUMMIT participants who only underwent single timepoint CT imaging, where baseline PPFE only associated with respiratory admissions but not dyspnoea severity or steroid/antibiotic use.

Both baseline and progressive PPFE were associated with increased respiratory admissions, higher mMRC scores, and greater use of steroid or antibiotic medications. These findings indicate that PPFE progression may reflect underlying inflammatory or fibrotic processes that increase vulnerability to chest infections. The borderline association between progressive PPFE and mMRC score suggests a biologically plausible link between progressive PPFE and worsening breathlessness, which may become statistically significant with longer follow-up.

The presence of ILAs showed comparable associations with respiratory hospitalizations and mMRC scores, but did not show associations with increased steroid/antibiotic use. This suggests that the clinical course of disease for PPFE might differ to that of classical UIP-type fibrotic damage. The 55\% increased incidence rate of steroid/antibiotic use in participants with progressive PPFE adds confidence that true excess respiratory morbidity is being captured. Yet within our data, overlap between PPFE and other interstitial processes cannot be entirely excluded. We also demonstrated that whilst several chronic lung diseases have been linked to systemic inflammation and increased cardiovascular morbidity, in screening populations, progressive PPFE reflects predominantly respiratory morbidity rather than heralding systemic cardiovascular disease risk. 

The reported results in this study have important clinical implications as lung cancer screening becomes more widely implemented. Radiologists and clinicians will increasingly identify incidental disease processes within and outside the lung which may contribute to morbidity or mortality in an individual. Improving recognition of these abnormalities could help modify the respiratory health of the vulnerable, high disease-burden population that forms the bedrock of lung cancer screening initiatives. Confident quantitation of progressive disease phenotypes, such as PPFE could also act as a source of recruitment into clinical trials of new therapies. The opportunity to run clinical trials in rare chronic lung diseases is likely to exponentially expand in the coming years with the roll-out of lung cancer screening programs and could transform the level of disease severity at which therapeutic intervention is considered for a range of lung conditions. Specifically with regard to PPFE, longitudinal quantitative PPFE assessment could enhance risk stratification in screening programmes, potentially informing follow-up imaging intervals, respiratory function monitoring, or early referral to ILD specialists. Furthermore, because progressive PPFE appears specific to respiratory outcomes and not cardiovascular events, it may serve as a targeted biomarker of respiratory risk in high-risk smokers.

There were several limitations to our study. First, the study design is observational and retrospective, and therefore does not permit causal inference. Although progressive PPFE was independently associated with mortality and respiratory morbidity, the mechanistic pathway linking radiologic PPFE progression to adverse outcomes remains incompletely defined. It is unclear which underlying pathological processes drive progression of PPFE-like lesions in lung cancer screening populations, or how such progression directly contributes to increased mortality risk. These findings should therefore be interpreted as observational associations rather than evidence of causality. Second, LDCT resolution is inherently lower than diagnostic HRCT, and although PPFE is typically high density, subtle lesions may have been underestimated. Third, PPFE and ILA may overlap anatomically, and despite careful visual adjudication and automated segmentation restricted to upper lobes, some degree of misclassification is possible, although our thorough visual review suggests this is likely minimal. Fourth, the definition of progressive PPFE was based on a distribution-derived threshold. While aligned with established biomarker methodology, external validation is required to confirm whether this threshold corresponds to clinically meaningful differences in other screening and non-screening populations. And lastly, we excluded participants without a second CT scan because longitudinal imaging was required to quantify progressive PPFE. This may introduce selection bias, as individuals with more advanced disease at baseline may not have survived long enough to undergo repeat imaging. To assess the potential impact of this exclusion, we performed a secondary baseline-only analysis in the SUMMIT cohort. In this subgroup, baseline PPFE remained independently associated with respiratory admissions, suggesting that PPFE burden identified at a single timepoint still captures clinically relevant respiratory risk.

In summary, radiologic PPFE progression has been shown to be a robust, reproducible predictor of respiratory morbidity and mortality in two large LCS cohorts. Quantitative assessment of progressive PPFE offers a sensitive, clinically relevant imaging biomarker that may enhance risk prediction and provide insight into early non-UIP fibrotic processes detectable within population-based screening programmes.

\section{Supplementary Materials}
\subsection*{Visual CT Evaluation of ILAs} 
The SUMMIT CT scans were visually evaluated for ILA by two expert thoracic radiologists (JJ, 16 years’ experience; DY, 10 years’ experience). The dataset was divided into two subsets and reviewed independently by each radiologist. Scans were classified into four categories: No ILA, non-fibrotic ILA (NF-ILA), fibrotic ILA (F-ILA), or undiagnosed interstitial lung disease (U-ILD) \citep{cheng2026fibrotic}. ILA was defined as the presence of non-dependent ground-glass opacification (GGO) and/or reticulation affecting more than one lobe. Fibrotic involvement was assessed using a four-point lobar traction bronchiolectasis score, where traction bronchiolectasis was defined as dilated bronchioles within the peripheral 2\,cm of lung occurring in areas of reticulation and/or GGO. Based on these criteria, NF-ILA was defined as GGO and/or reticulation affecting two or more lobes without traction bronchiolectasis. Fibrotic ILA (F-ILA) met the same criteria, but with traction bronchiolectasis present in one or two lobes. Undiagnosed fibrotic ILD (U-ILD) was defined by the presence of traction bronchiolectasis in three or more lobes. The lingula segment of the left upper lobe was considered a separate lobe. All scans identified as ILA by one reviewer, together with a random sample of 10\% of scans classified as No ILA, were re-evaluated by the second reviewer in a blinded manner. Cases with disagreement were resolved through consensus review. For downstream analyses in this study, the visual categories were converted into a binary ILA variable to ensure consistent representation across the dataset: scans labelled as No ILA were coded as 0, whereas scans labelled as NF-ILA, F-ILA, or U-ILD were coded as 1. Scans labelled as exclusion were treated as missing values. A detailed description of the visual assessment protocol is available in \citep{cheng2026fibrotic}.

\begin{figure*}
\centering
\includegraphics[width=\linewidth]{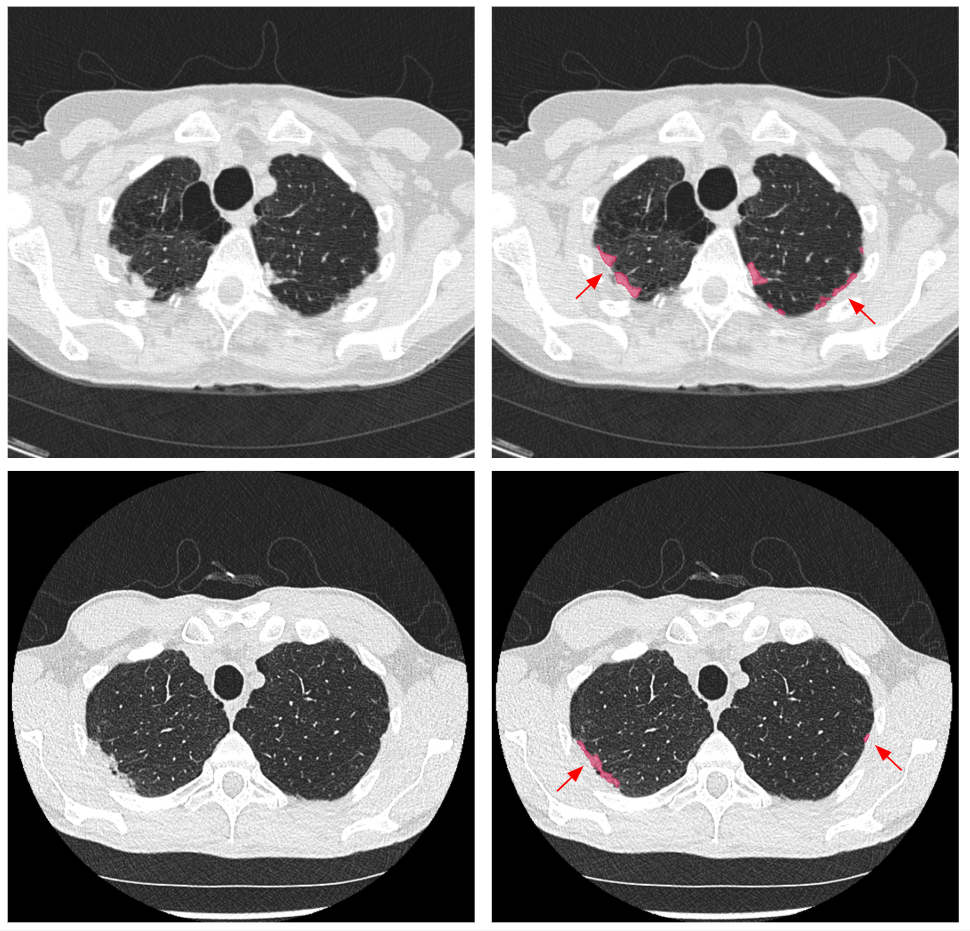}
\caption{Representative examples of automated PPFE quantification on axial CT images from the NLST and SUMMIT cohorts. The first column shows the original CT images, and the second column shows the corresponding CT images with automated PPFE segmentation overlaid. Highlighted regions (red) indicate PPFE identified by the automated segmentation model in the upper lung zones. The first row shows images from a 65-year-old female participant in the NLST cohort with a quantified PPFE volume of 8.85~cm$^{3}$. The second row shows images from a 55-year-old male participant in the SUMMIT cohort with a quantified PPFE volume of 5.32~cm$^{3}$.}
\label{Segmentation_example}
\end{figure*}

\subsection*{Automated CT-based PPFE Quantification}
An automated deep learning-based model was developed to segment PPFE lesions across all available CT scans. The model architecture was based on a U-Net framework~\cite{ronneberger2015u,isensee2021nnu}, consisting of an encoder-decoder structure with skip connections to preserve spatial information during upsampling. PPFE lesions were manually identified and segmented in 100 CT scans from the SUMMIT dataset to provide training annotations. Only lesions located in the upper lung zones (above the carina, approximately corresponding to the upper lobes) were considered. PPFE-like lesions were defined radiologically as dense, pleural-based or subpleural consolidative opacities predominantly involving the upper lobes, typically with triangular, wedge-shaped, or band-like morphology, and associated with subjacent architectural distortion or volume loss. Isolated thin apical caps confined to the most cranial pleural margin without associated subpleural fibrosis were not considered PPFE. We utilised two strategies to minimise inclusion of physiological apical pleural thickening and apical cap fibrosis—which may be present in healthy individuals or in non-PPFE lung disease. Firstly the most apical 5\,mm of the lung parenchyma was excluded from analysis. Second, we focused on change in PPFE over a 2-year interval, which should not occur in apical pleural thickening or benign apical fibrosis.

Model training was performed using 5-fold cross-validation on the annotated dataset, with 80\% of cases used for training and 20\% held out for testing in each fold. Training was conducted for 1000 epochs using stochastic gradient descent. The loss function consisted of a compound Dice and cross-entropy loss, enabling robust learning under class imbalance. Segmentation performance was evaluated using the Dice similarity coefficient (DSC)~\cite{milletari2016v}, which quantifies spatial overlap between predicted and manual segmentations. The model achieving the highest mean DSC across folds was selected as the final model and applied for inference across all remaining CT scans. Across the five folds, the selected model achieved a mean DSC of 0.91, indicating robust agreement with expert annotations.

For each participant, PPFE lesions were automatically segmented on baseline and follow-up low-dose CT scans. PPFE volume was quantified by multiplying the number of voxels classified as PPFE by the voxel dimensions, yielding a total volume in cubic centimetres (cm$^3$) for each scan. These quantitative PPFE measures were subsequently used to derive longitudinal PPFE change metrics for downstream statistical analyses. An example of automated PPFE segmentation is shown in Figure~\ref{Segmentation_example}.

\begin{figure}
\centering
\includegraphics[width=0.6\linewidth]{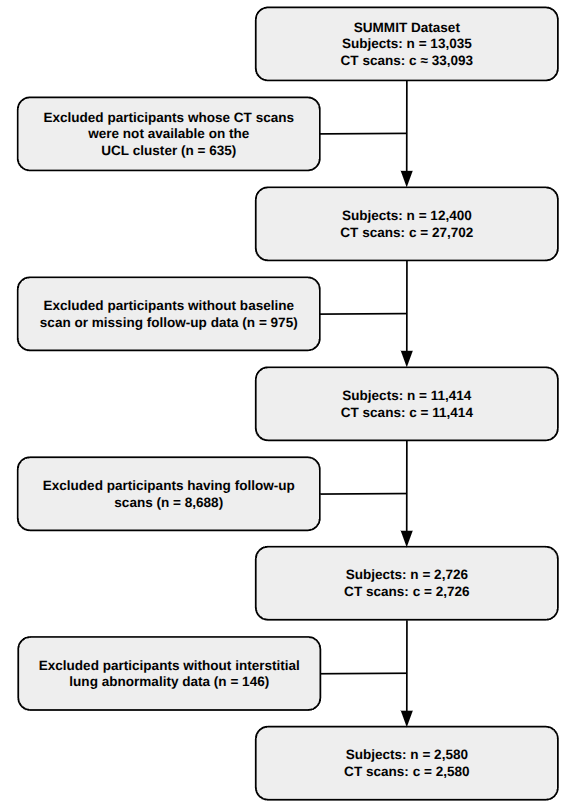}
\caption{Flowchart illustrating identification of baseline-only participants in SUMMIT cohort. 
This diagram outlines the selection of SUMMIT participants who did not undergo a follow-up CT scan. Starting from the full SUMMIT cohort (n=13,035), individuals with available baseline CT imaging were identified, after which participants with follow-up CTs were excluded. Additional exclusions were applied based on imaging quality and availability of clinical data. The final baseline-only subgroup consisted of 2,580 participants.}
\label{flowchart_baseline}
\end{figure}

\subsection*{Baseline-only Analysis on SUMMIT Cohort}
To further investigate the prognostic significance of radiologic PPFE within the SUMMIT cohort, we performed an additional sub-analysis focused on participants who did not return for a follow-up CT scan during the screening (baseline-only). These individuals were identified after excluding all participants with complete longitudinal imaging. The absence of follow-up imaging frequently reflected clinical or logistical factors, including severe comorbidity, frailty, or health deterioration that prevented return for routine screening. This subgroup therefore provided an opportunity to evaluate whether baseline PPFE alone was clinically informative in individuals who lacked follow-up CT scans. Specifically, we assessed whether participants with only a baseline scan demonstrated PPFE patterns that might still relate to adverse clinical outcomes. 

The workflow used to identify the baseline-only subgroup is shown in Supplementary Figure~\ref{flowchart_baseline}. After applying identical inclusion criteria for imaging quality and clinical data completeness, a total of 2,580 SUMMIT participants were identified. Baseline PPFE was quantified using the same automated deep-learning segmentation model applied in the main analysis. For clinical interpretability, baseline PPFE volume was categorised using the pre-defined PPFE progression threshold described in the main manuscript, namely by using half the standard deviation of baseline PPFE in NLST. Although the threshold originally designed to represent a minimal clinically important PPFE progression, this threshold was used here to denote meaningful baseline PPFE burden. Participants with baseline PPFE volumes above this threshold were classified as having clinically important PPFE, whereas those with volumes below the threshold were classified as having non-clinically important PPFE. Using above-mentioned approach, the baseline-only cohort was divided into 1,733 participants with clinically important PPFE and 847 with non-clinically important PPFE. These subgroups are summarised in Supplementary Table~\ref{tab_demographics_baseline}, which includes demographic characteristics, smoking history, FVC \% predicted, height, visual ILA status, and baseline PPFE. This table follows the same structure as the main demographic table (Table~\ref{tab_demographics_change}), enabling direct comparison with participants who completed longitudinal imaging.

To evaluate whether baseline PPFE burden alone was associated with adverse respiratory or clinical outcomes, we repeated the main modelling framework in the baseline-only cohort. Longitudinal PPFE metrics were not available for these participants; therefore, all models included baseline PPFE as the primary predictor. Respiratory hospital admission counts were analysed using Negative Binomial generalised linear models to account for overdispersed count data. Ordinal logistic regression was used to assess the association between baseline PPFE and dyspnoea severity, measured using the mMRC score. Use of respiratory-related anti-inflammatory medications -antibiotics or steroid- was analysed using Negative Binomial regression, with cumulative prescription counts as the dependent variable. All models were adjusted for age, sex, smoking history (pack-years), FVC \% predicted, and ILA. Results were expressed as incidence rate ratios (IRRs) or odds ratios (ORs) with 95\% confidence intervals. All supplementary analyses were performed using Python (version 3.11) with the lifelines, statsmodels, and scikit-survival libraries.

Within the SUMMIT cohort, 2,580 participants only received a baseline CT and no follow-up CT scan. Supplementary Table~\ref{tab_demographics_baseline} summarises their demographic and clinical characteristics. This cohort was older than the main SUMMIT cohort (Table~\ref{tab_demographics_change} - median age 65 years), had a similar sex distribution (42\% female), and smoking history when compared to participants who had a 2-year follow-up scan. The subset of subjects with clinically important PPFE (defined using the pre-specified PPFE threshold from the main analysis) had much higher baseline PPFE volumes (mean 1.62~cm\textsuperscript{3}) compared with those with non-clinically important PPFE (mean 0.09~cm\textsuperscript{3}). The group with clinically important PPFE had a higher prevalence of visually detected ILA (23\% vs. 13.7\%), suggesting that even in the absence of longitudinal scans, increased PPFE may co-occur with other interstitial abnormalities. These descriptive findings are consistent with the main analysis, where baseline PPFE and progression were associated with respiratory outcomes and mortality.

Supplementary Table~\ref{tab_neg_binomial_respiratory_summit} presents the negative binomial regression results examining respiratory admissions in the baseline-only subgroup. Older age, increased smoking history (pack-years), 
lower FVC~\% predicted, and the presence of ILA were all associated with higher admission 
rates. Baseline PPFE demonstrated a strong and independent association with respiratory admissions (IRR~1.24, 95\% CI~1.17--1.31, $\text{p}\raisebox{0.15ex}{$<$}0.001$), mirroring the main analysis where both baseline PPFE and $\Delta$PPFE predicted respiratory admissions. 

Supplementary Table~\ref{tab_ordinal_logistic_summit_y0} shows the ordinal logistic regression results for the mMRC score. Older age, male sex, increased smoking history (pack-years), and lower FVC were significant predictors 
of worse dyspnoea. In contrast with the main analysis, baseline PPFE did not show a significant association with the mMRC score (OR~1.01, p=0.701). 

Supplementary Table~\ref{tab_neg_binomial_steroid_y1_summit} reports the negative binomial model for steroid/antibiotic use. Predictors of increased medication usage included older age and reduced FVC~\% predicted. 
However, baseline PPFE was not associated with increased medication use (IRR~0.96, p=0.436). In the main analysis, progressive PPFE was the primary determinant of medication use. 

The above-described supplementary analyses suggests that Baseline PPFE in participants with single scans showed a robust association with respiratory admissions, reinforcing PPFE as an independent marker of respiratory morbidity. Lack of association with mMRC or steroid/antibiotic medication use suggests that PPFE progression is a more sensitive indicator of symptoms and acute respiratory events. Moreover, higher ILA prevalence among clinically important PPFE individuals suggests overlap in structural lung abnormalities, consistent with findings in the main cohort. Therefore, the results suggest that PPFE even when measured at a single time point carries clinically relevant information. However, the strongest prognostic value in the main study derived from progressive PPFE, highlighting the importance of longitudinal imaging when available.

\begin{table*}[htbp]
\centering
\caption{Overall cohort characteristics in NLST and SUMMIT. Variables include age, sex, height, smoking history (pack-years), FVC, visual scores of ILA, and computer-based scores of PPFE. PPFE (baseline) = upper-lung pleuroparenchymal fibroelastosis quantified on the first scan, PPFE (follow-up) = upper-lung pleuroparenchymal fibroelastosis quantified on the follow-up scan, ILA = visual scores of interstitial lung abnormality, FVC = forced vital capacity. PPFE (baseline), PPFE (follow-up), Height, and FVC are described as Mean$\pm$SD. Age and smoking are described as median (range). Sex is described as Female/Male \%. ILA is described as number/\%.}
\begingroup
\scriptsize
\setlength{\tabcolsep}{4pt}
\renewcommand{\arraystretch}{1.25}
\begin{tabular}{lcc}
\toprule
\textbf{Variable} & \textbf{NLST} & \textbf{SUMMIT} \\
\midrule

Num Patients, (\%) 
& 7980 (100) & 8561 (100) \\

Age y, median (range)
& 61 (54--74) & 65 (55--88) \\

Sex (F/M\%)
& 41/59 & 42/58 \\

Smoking py, median (range)
& 48 (21--272) & 40 (2.5--400) \\

FVC \%pred, mean$\pm$SD
& -- & 88.07$\pm$16.53 \\

Height cm, mean$\pm$SD
& 172.44$\pm$9.98 & 168.93$\pm$9.79 \\

ILA, n(\%)
& -- & 1358 (15.8) \\

PPFE (baseline scan), mean$\pm$SD 
& 0.35$\pm$0.81 & 0.52$\pm$1.05 \\

PPFE (follow-up scan), mean$\pm$SD 
& 0.35$\pm$0.80 & 0.56$\pm$1.08 \\

\bottomrule
\end{tabular}
\par\smallskip
\begin{flushleft}
\end{flushleft}
\endgroup
\label{tab_cohort_summary_all}
\end{table*}

\begin{table*}[htbp]
\centering
\caption{Patient demographics and clinical variables in the SUMMIT cohort comparing clinically important PPFE (C-PPFE) vs non-clinically important PPFE (NC-PPFE). Baseline PPFE refers to upper-lung pleuroparenchymal fibroelastosis quantified on the baseline scan. Baseline PPFE, Height, and FVC are reported as mean$\pm$SD. Age and smoking are medians (range). Sex is Female/Male\%. ILA is n(\%).}

\begingroup
\scriptsize
\setlength{\tabcolsep}{4pt}
\renewcommand{\arraystretch}{1.35}
\begin{tabular}{lccc}
\toprule
\textbf{Variable} &
\textbf{Non-clinically Important PPFE} &
\textbf{Clinically Important PPFE*} &
\textbf{P-value} \\
\midrule

Num Patients &
1733 & 847 & -- \\

Age y, median (range) &
65 (55--78) & 67 (55--78) & $\raisebox{0.15ex}{$<$}0.001$ \\

Sex, F/M\% &
42/58 & 42/58 & 0.041 \\

Smoking py, median (range) &
65 (55--78) & 43 (8--252) & 0.024 \\

FVC \%pred, mean$\pm$SD &
84.35$\pm$17.12 & 85.18$\pm$16.79 & 0.242 \\

Height cm, mean$\pm$SD &
167.68$\pm$9.09 & 169.82$\pm$10.53 & $\raisebox{0.15ex}{$<$}0.001$ \\

ILA, n(\%) &
239 (13.7) & 196 (23.0) & $\raisebox{0.15ex}{$<$}0.001$ \\

Baseline PPFE, mean$\pm$SD &
0.09$\pm$0.10 & 1.62$\pm$1.54 & $\raisebox{0.15ex}{$<$}0.001$ \\

\bottomrule
\end{tabular}

\par\smallskip
\begin{flushleft}
\scriptsize Clinically Important PPFE* = PPFE $\geq$ 0.41 cm$^3$.
\end{flushleft}

\endgroup
\label{tab_demographics_baseline}
\end{table*}

\begin{table*}[htbp]
\centering
\caption{Negative Binomial GLM model for respiratory admissions in the SUMMIT cohort. Model adjusted for age, sex, smoking (pack-years), FVC \% predicted, and ILA. IRRs in bold indicate p < 0.05. Baseline PPFE: upper-lung pleuroparenchymal fibroelastosis quantified on the first scan, ILA: visual scores of interstitial lung abnormality, FVC: forced vital capacity.}
\begingroup
\scriptsize
\setlength{\tabcolsep}{6pt}
\renewcommand{\arraystretch}{1.12}
\begin{tabular}{lccc}
\toprule
\textbf{Variable} & \textbf{Incidence Rate Ratio} & \textbf{95\% confidence interval} & \textbf{p-value} \\
\midrule
Age (y)             & \textbf{1.060} & 1.042--1.078 & $\raisebox{0.15ex}{$<$}0.001$ \\
Sex (male)                & 0.847          & 0.686--1.045 & 0.121  \\
Smoking (pack-years)    & \textbf{1.005} & 1.002--1.009 & $\raisebox{0.15ex}{$<$}0.001$  \\
FVC \% predicted        & \textbf{0.976} & 0.970--0.982 & $\raisebox{0.15ex}{$<$}0.001$ \\
ILA (present vs absent)   & \textbf{1.546} & 1.208--1.980 & $\raisebox{0.15ex}{$<$}0.001$  \\
Baseline PPFE   & \textbf{1.238} & 1.167--1.314 & $\raisebox{0.15ex}{$<$}0.001$ \\
\bottomrule
\end{tabular}
\par\smallskip
\caption*{}
\endgroup
\label{tab_neg_binomial_respiratory_summit}
\end{table*}

\begin{table*}[htbp]
\centering
\caption{Ordinal logistic regression model for mMRC score in the baseline-only SUMMIT cohort. Model adjusted for age, sex, smoking (pack-years), FVC \% predicted, and ILA. Odds in bold indicate p < 0.05. Baseline PPFE: upper-lung pleuroparenchymal fibroelastosis quantified on the first scan, ILA: visual scores of interstitial lung abnormality, FVC: forced vital capacity.}
\begingroup
\scriptsize
\setlength{\tabcolsep}{6pt}
\renewcommand{\arraystretch}{1.12}
\begin{tabular}{lccc}
\toprule
\textbf{Variable} & \textbf{Odds Ratio} & \textbf{95\% confidence interval} & \textbf{p-value} \\
\midrule
Age (y)                     & \textbf{1.019} & 1.007--1.031 & 0.002 \\
Sex (male)               & \textbf{0.649} & 0.559--0.753 & $\raisebox{0.15ex}{$<$}0.001$ \\
Smoking (py)                & \textbf{1.011} & 1.008--1.015 & $\raisebox{0.15ex}{$<$}0.001$ \\
FVC \% predicted            & \textbf{0.973} & 0.969--0.977 & $\raisebox{0.15ex}{$<$}0.001$ \\
ILA (present vs absent)     & 1.084          & 0.891--1.319 & 0.419 \\
Baseline PPFE               & 1.012 & 0.951--1.077 & 0.701 \\
\bottomrule
\end{tabular}
\par\smallskip
\caption*{}
\endgroup
\label{tab_ordinal_logistic_summit_y0}
\end{table*}

\begin{table*}[htbp]
\centering
\caption{Negative Binomial GLM model for steroid/antibiotic inflammatory medication use in the baseline-only SUMMIT cohort. Model adjusted for age, sex, smoking (pack-years), FVC \% predicted, and ILA. IRRs in bold indicate p < 0.05. Baseline PPFE: upper-lung pleuroparenchymal fibroelastosis quantified on the first scan, ILA: visual scores of interstitial lung abnormality, FVC: forced vital capacity.}
\begingroup
\scriptsize
\setlength{\tabcolsep}{6pt}
\renewcommand{\arraystretch}{1.12}
\begin{tabular}{lccc}
\toprule
\textbf{Variable} & \textbf{IRR} & \textbf{95\% confidence interval} & \textbf{p-value} \\
\midrule
Age (y)                  & \textbf{1.024} & 1.006--1.042 & 0.007 \\
Sex (male)            & 0.874          & 0.703--1.088 & 0.229 \\
Smoking (py)             & 0.995          & 0.990--1.001 & 0.085 \\
FVC \% predicted         & \textbf{0.987} & 0.981--0.993 & $\raisebox{0.15ex}{$<$}0.001$ \\
ILA (present vs absent)  & 0.848          & 0.617--1.166 & 0.312 \\
Baseline PPFE            & 0.963        & 0.878--1.057 & 0.436 \\
\bottomrule
\end{tabular}
\par\smallskip
\caption*{}
\endgroup
\label{tab_neg_binomial_steroid_y1_summit}
\end{table*}

\section*{Acknowledgments}
SA and this research was supported by the International Alliance for Cancer Early Detection, an alliance between Cancer Research UK [EDDAPA-2023/100002], Canary Center at Stanford University, the University of Cambridge, OHSU Knight Cancer Institute, University College London and the University of Manchester. JJ and this work was funded in whole or in part by the Wellcome Trust (227835/Z/23/Z). JJ and SMJ were also supported by the NIHR UCLH Biomedical Research Centre, UK. The SUMMIT Study is funded by GRAIL LLC. through a research grant awarded to SMJ as Principal Investigator. DY is supported by grants from the fellowship of Astellas Foundation for Research on Metabolic Disorders. DC is supported by a Clinical Research Training Fellowship from the Alliance for Cancer Early Detection (ACEDAS-2023/100003). The authors thank the National Cancer Institute for access to NCI’s data collected by National Lung Screening Trial (NLST). The authors would like to thank the family of Nathalie Soffe for their support of Breathing Matters (a UCLH Charity) and for helping to further research into improving the understanding of PPFE development. The statements contained herein are solely those of the authors and do not represent or imply concurrence or endorsement by NCI.

\noindent\textbf{SUMMIT Consortium:} Aashna Samson$^{4}$, Adefiola Olabode$^{4}$, Alberto Villanueva$^{4}$, Ali Mohammed$^{13}$, Alice Cotton$^{4}$, Andrew Crossingham$^{4}$, Andrew Perugia$^{22}$, Anjeli Ketkar$^{4}$, Anna Sikorska$^{4}$, Anthony Edey$^{4}$
% ...

\noindent\textbf{Affiliations:} 1- Lungs For Living Research Centre, UCL Respiratory, University College London, London; 2- CRUK \& UCL Cancer Trials Centre, University College London, London; 3- Centre for Cancer Screening, Prevention, Detection and Early Diagnosis, Wolfson Institute of Population Health, Barts and The London School of Medicine and Dentistry, Queen Mary University of London, London; 4- University College London Hospitals NHS Foundation Trust, London; 5- Royal Brompton and Harefield NHS Foundation Trust, London; 6- National Heart and Lung Institute, Imperial College, London; 7- Royal Free London NHS Foundation Trust, London; 8-Whittington Health NHS Trust, London; 9- Barking, Havering and Redbridge University Hospitals NHS Trust, Essex; 10- Homerton University Hospital Foundation Trust, London; 11- The Princess Alexandra Hospital NHS Trust, Essex; 12- North Middlesex University Hospital NHS Trust, London; 13- Barts Health NHS Trust, London; 14- North Bristol NHS Trust, Bristol; 15- Royal United Hospitals Bath NHS Foundation Trust, Bath; 16- Surrey and Sussex Healthcare NHS Trust, Surrey; 17- King's College Hospital NHS Foundation Trust, London; 18- University Hospitals Sussex NHS Foundation Trust, Sussex; 19- Royal Papworth Hospital NHS Foundation Trust, Cambridge; 20-Satsuma Lab, Hawkes Institute, London; 21-Tower Hamlets Clinical Commissioning Group, London; 22- Noclor Research Support, London; 23- Islington Clinical Commissioning Group, London.

\section*{Conflicts of Interest}
JJ declares consultancy fees from Boehringer Ingelheim, F. Hoffmann-La Roche, GlaxoSmithKline, NHSX; Advisory Boards for Boehringer Ingelheim, F. Hoffmann-La Roche;Lecture fees from Boehringer Ingelheim, F.Hoffmann-La Roche, Takeda; Grant Funding: from GlaxoSmithKline, Wellcome Trust, Microsoft Research, Gilead Sciences; Patents: UK patent application numbers 2113765.8 and GB2211487.0. SMJ is supported by CRUK programme grant (EDDCPGM/100002), and MRC Programme grant (MR/W025051/1). SMJ receives support from the CRUK Lung Cancer Centre of Excellence (C11496/ A30025) and the CRUK City of London Centre, the Rosetrees Trust, the Roy Castle Lung Cancer foundation, the Longfonds BREATH Consortia, MRC UKRMP2 Consortia, the Garfield Weston Trust and University College London Hospitals Charitable Foundation. SMJ’s work is supported by a Stand Up To Cancer-LUNGevity- American Lung Association Lung Cancer Interception Dream Team Translational Research Grant and Johnson and Johnson (grant number: SU2C-AACR-DT23-17 to S.M. Dubinett and A.E. Spira). Stand Up To Cancer is a division of the Entertainment Industry Foundation. SMJ has received fees for advisory board membership in the last three years from Bard1 Lifescience. He has received grant income from GRAIL Inc. He is an unpaid member of a GRAIL advisory board. He has received lecture fees for academic meetings from Cheisi and Astra Zeneca. His wife works for Astra Zeneca. DCA is Director and a shareholder of Queen Square Analytics. ACPS received honoraria from Roche and educational fees from Amgenreceived honoraria from Roche and educational fees from Amgen.

\section*{}
\bibliographystyle{elsarticle-harv}
\bibliography{ref.bib}
\end{document}